
\magnification=\magstep1
\hsize=5.5 true in
\centerline{\bf THE STRUCTURE OF NAKED SINGULARITY IN SELF-SIMILAR }
\centerline{\bf GRAVITATIONAL  COLLAPSE II}
\vfill
\centerline{\bf P. S. Joshi and I. H. Dwivedi$^{*}$ }
\centerline{\bf Tata Institute of Fundamental Research}
\centerline{\bf Homi Bhabha Road, Bombay 400 005}
\centerline{\bf India}
\vfill
\noindent {\bf $^{*}$ Permanent Address}:\hfill\break
\noindent{\bf Institute of Basic Sciences}\hfill\break
\noindent{\bf Agra University}\hfill\break
\noindent{\bf Khandari, Agra, India}\hfill\break
\vfill
\noindent{\bf Proofs to be sent to:}\hfill\break
\noindent{\bf Dr. Pankaj  S. Joshi}\hfill\break
\noindent{\bf Theoretical Astrophysics Group}\hfill\break
\noindent{\bf T.I.F.R. Homi Bhabha Road}\hfill\break
\noindent{\bf Colaba, Bombay 400 005}\hfill\break
\noindent{\bf India}\hfill\break
\vfill

\vfil\eject

\magnification=\magstep1  
\hoffset=0 true cm        
\hsize=6.0 true in        
\vsize=8.5 true in        
\baselineskip=24 true pt plus 0.1 pt minus 0.1 pt 
   e spacing) 72.27 pt=1 inch
\overfullrule=0pt         

\beginsection {Abstract}

Generalizing the results of Joshi and Dwivedi in Commun.Math.Phys.
146, p.333 (1992), it is pointed out that  strong curvature naked
singularities  could occur in the self-similar gravitational collapse
of any form of matter satisfying the weak energy condition for the
positivity of mass-energy density.

\vfil\eject

The  formation of naked singularities  in self-similar
gravitational collapse of a perfect fluid with an adiabatic equation
of state was analyzed in [1] (to be
referred to here as
paper I). It was shown that a powerfully strong curvature
naked singularity forms in that case from which a non-zero measure set of
non-spacelike trajectories come  out, which could be locally or
globally naked. As there is no proof
available as yet for the cosmic censorship hypothesis, which is fundamental
to the black hole physics and its applications,
such an analysis of collapse
scenarios appears inevitable in order to prove, and even to formulate a
mathematically rigorous  censorship hypothesis. This would
allow the features of collapse to be isolated which must be avoided
in order to establish cosmic censorship.

Eventhough the form of matter such as a perfect fluid has a
wide range of physical applications with the advantage of incorporating
the pressure which could be important in the later stages of collapse,
it is certainly important to examine if similar conclusions will hold
for  other reasonable forms of matter.
In fact, as pointed out by
Eardley [2], the forms of matter such as dust could be an approximation to
more fundamental forms, such as a massive scalar field.
It is thus conceivable that the naked singularity could be an
artifact of the approximation used, rather than being a basic feature of
gravitational collapse, and therefore
it is important to examine these conclusions for a broader range of matter.

With this purpose, we examine  here the final fate of a self-similar
collapse to show
that  conclusions on the occurrence and nature of naked singularity
generalize to all forms of matter which satisfy the positivity of
energy as characterized by the well-known energy conditions [3].
It follows that the conclusions of paper I are not limited to a
specific form of matter such as a perfect fluid.

The non-zero metric components for a spherically symmetric space-time are
($t,r,\theta,\phi= 0,1,2,3$)
$$g_{00}=-e^{2\nu},\quad g_{11}=e^{2\psi}\equiv V+X^2e^{2\nu},\quad g_{33}=
g_{22}\sin ^2\theta =r^2S^2 \sin ^2\theta \eqno(1)$$
where $V$ is defined as above and due to self-similarity
$\nu,\psi, V$
and $S$ are functions of the similarity parameter $X=t/r$ only.
The remaining freedom in the choice
of coordinates $r$ and $t$ has been used to set the only off-diagonal term
$T_{01}$ of the energy momentum
tensor $T_{ij}$ to zero (using comoving coordinates).
We assume the matter to satisfy the weak energy condition, i.e.
$$T_{ij}V^iV^j\ge 0\eqno(2)$$
for all non-spacelike vectors $V^i$.
Therefore, ${T_0}^{0}\le 0$
( i.e $T_{00}\ge 0$), $
{T_1}^{1}-{T_0}^{0}\ge 0$, ${T_2}^{2}-{T_0}^{ 0}\ge 0$.
The relevant field
equations for a spherically symmetric self-similar collapse of the
fluid under consideration are [4],
$${G^0}_0={-1\over S^2}+{2e^{-2\psi}\over S}(X^2\ddot S -X^2\dot S\dot\psi
+XS\dot \psi+{(S-X\dot S)^2\over 2S})
-{2e^{-2\nu}\over S}(\dot S\dot\psi +{ \dot S^2\over 2S})
=8\pi r^2{T^0}_0 \eqno(3a)$$
$${G^1}_1={-1\over S^2}-{2e^{-2\nu}\over S}(\ddot S-\dot S\dot\nu+
{\dot S^2\over 2S}) +{2e^{-2\psi}\over S}[-SX\dot \nu+X^2\dot S\dot\nu+{(S
-X\dot S)^2 \over 2S}]=8\pi r^2{T^1}_1 \eqno(3b)$$
$${G^2}_2=8\pi T_2^2,\quad
{G^0}_1=\ddot S-\dot S\dot \nu-\dot S\dot\psi+{S\dot \psi\over X}={T^0}_1
=0\eqno(3c)$$
Using equation (3c), equations (3a) and (3b)
can be combined to get
$$\dot V(X)=
Xe^{2\nu}[H-2]\eqno(3d)$$
where (${\cdot}$) is the derivative with respect to the similarity parameter
$X=t/r$ and $H=H(X)$ is defined by
$$H =r^2e^{2\psi}
({T^1}_1-{T_0}^{0})\eqno(4)$$
For matter satisfying weak energy
condition, it follows that $H(X)\ge 0$ for all $X$.
Here we have four field equations with six unknowns. The remaining
two equations come from the choice of form of matter one is dealing with.
Using (3b) and methods similar to paper I one can see that the
singularity at $t=0,r=0$ is naked
when the equation $V(X)=0$ has a real simple
positive root, i.e.
$$V(X_0)=0\eqno(5)$$
for some $X=X_0$; (see also Waugh and Lake [6], and the use of such a
condition in the numerical simulations by Ori and Piran [6]; and also [5]
for the relevance of the same for naked singularity formation in self-similar
radiation collapse models); and a non-zero measure
of future directed non-spacelike trajectories will escape
from the singularity provided
$$0<H_0=H(X_0)<\infty \eqno(6)$$

These escaping trajectories near the singularity are given by,
$$r= D(X-X_0)^{2\over H_0-2}\eqno(7)$$
Here $D$ is a constant labeling different integral curves, which are the
solutions of the geodesic equations, coming out of the
naked singularity.
It is seen that $H_0>0$ will hold if the weak energy condition (2) is
satisfied and when the energy density as measured by any timelike observer
is positive in  the collapsing region near the singularity.
In this case, when $H_0<\infty$, families of future directed
non-spacelike geodesics will come out, terminating at
the naked singularity in the past. On the other hand, for $H_0=\infty$, a
single non-spacelike trajectory will come out of the naked singularity.
This characterizes
the  formation of  naked singularity in  self-similar gravitational collapse.
Such a singularity will be atleast locally naked
and considerations such as those in paper I can be used to show that
it could be globally naked as well
provided $V(X_0)=0$ has more than one real
simple positive roots.

We do not go here into a detailed discussion of the sufficient
conditions which would ensure the real positive roots for (5)
(and hence the existence of naked singularity) in terms of physical
parameters involved and the initial data specified prior to the onset of
the singularity.
However, the existence of several classes of self-similar
solutions to Einstein equations, where the gravitational collapse
from a regular initial data results
into a naked
singularity, indicates that such a condition will be realized for a wide
variety of self-similar collapse scenarios.  For example, the cases
of radiation collapse models with a linear mass function
[5], and self-similar Tolman-Bondi models [6] are all
special cases of the treatment given here. For the case of
radiation collapse with  the mass function $m(u)=\lambda u$
(where $u$ is the advanced time), the above initial condition corresponds
to a restriction on the parameter
$\lambda$ (which is the rate of collapse)
given by $0<\lambda\le 1/8$. The collapse
must result into a naked singularity when $\lambda$ is in this range, and an
event horizon covers the singularity otherwise. Similar restrictions
are obtained, ensuring the existence of a real positive root for $V(X)=0$,
for classes of Tolman-Bondi dust models [7] and adiabatic perfect fluid [1]
( in which case this turns out to be a forth order
algebraic equation).  In general, for a given form of matter, the existence
of a real positive root of (5) would put a restriction on the range of values
of physical parameters involved.
One could also treat  condition (5) for the existence of a naked
singularity
as an initial value problem for the differential equation (3d).
That is, for a given form of matter
one could solve this first order differential equation  governing
$V(X)$,
with the initial value $V(X_0)=0$, for some
real positive value  $X=X_0$. It is possible
that the regular initial data specified for
a realistic gravitational collapse might just always avoid this initial
condition and could result into a black hole. On the other hand, all the
initial  data sets satisfying
this condition will necessarily evolve into
a naked singularity regardless of the form of matter involved.
It is thus seen that a naked singularity would form
in the collapse of a wide range of matter forms satisfying the weak
energy condition.

Apart from non-zero measure of non-spacelike families coming out, the
curvature strength of  naked singularity provides an important test of
its physical significance. This
was calculated in paper I along all the non-spacelike geodesics terminating
at the naked singularity in the past to show that this is a strong curvature
singularity in a powerful sense. Even for the case of the general form of
matter
considered here, this turns out to be a strong curvature naked
singularity. We consider here only the radial null geodesics coming out
which are given by,
$${dt\over dr}= e^{\psi-\nu}\eqno(8)$$
The curvature strength of the singularity is measured by evaluating the
limit of $k^2 R_{ij}V^iV^j$ along these trajectories near the singularity,
which is given in the
present case by,
$$ \lim_{k\to 0} k^2R_{ij}V^iV^j= {4H_0\over (2+H_0)^2}> 0\eqno(9)$$

It follows that this is a strong curvature naked singularity in the sense
that the volume forms defined by all possible Jacobi vector fields
vanish in the limit of approach to the naked singularity, in which case
the space-time may not admit any continuous extension through
the singularity [8].

\vfil\eject

\centerline {$\underline {\cal REFERENCES}$}
\item{\bf 1.} P.S.Joshi and I.H.Dwivedi, Commun. Math. Phys. 146,333(1992).
\item{\bf 2.} D. M. Eardley in "Gravitation in Astrophysics", Plenum
Publishing Corporation Edited by B. Carter and J. B. Hartle 223 (1987).
\item{\bf 3.} S.W.Hawking and G.F.R.Ellis, "The Large Scale Structure of
Space-time", CUP, Cambridge (1973).
\item{\bf 4.} M.E. Cahill and A.HY.Taub, Commun.Math.Phys. 21, 1(1971).
\item{\bf 5.} I.H. Dwivedi and P.S. Joshi, Class. Quantum Grav.
6, 1599
(1989); 8, 1339 (1991).
\item{\bf 6} B.Waugh and K.Lake, Phys. Rev. D38, 4, 1315 (1988);
A.Ori and T.Piran, Phys. Rev. D42, 4, 1068 (1990).
\item{\bf 7} I.H.Dwivedi and P.S.Joshi, Class. Quantum Grav.
9, L69 (1992).
\item{\bf 8.} F.J.Tipler, C.J.S.Clarke and G.F.R.Ellis,
"General Relativity and Gravitation", Vol 2, (ed.A.Held),
New York: Plenum Press, 97 (1980).

\end